\documentclass[aps,prd,preprint,superscriptaddress,showpacs]{revtex4}
\usepackage{epsfig}
\usepackage{dcolumn}
\usepackage{bm}
\usepackage{amsmath}
\usepackage{amsfonts}
\usepackage{amssymb}
\usepackage{graphicx}
\usepackage{latexsym}
\usepackage{rotating}
\usepackage{multirow}
\usepackage{slashed}
\usepackage{xcolor}
\usepackage{hyperref}

\def\nnd{\end{document}}

\def\be{\begin{equation}}
\def\ee{\end{equation}}

\newcommand{\bea}{\begin{eqnarray}}
\newcommand{\eea}{\end{eqnarray}}
\newcommand{\bwt}{\begin{widetext}}
\newcommand{\ewt}{\end{widetext}}

\def\u
\def\hZ{\widehat Z}

\def\eed{\end{document}}

\def\m_z{m_{\textrm {Z}}}

\renewcommand{\u}{\rm{u}}

\def\be{\beta}

\def\rm#1{\textrm{#1}}

\makeatother
\begin{document}
\title{Natural Supersymmetry from the Yukawa Deflect Mediations}

\author{Tai-ran Liang}

\affiliation{School of Physics and Electronic Information, Inner Mongolia University for The Nationalities, Tongliao 028043, P. R. China}

\author{Bin Zhu}

\affiliation{Department of Physics, Yantai University, Yantai 264005, P. R. China}

\author{Ran Ding}

\affiliation{Center for High-Energy
Physics, Peking University, Beijing, 100871, P. R. China}

\author{Tianjun Li}

\affiliation{Key Laboratory of Theoretical Physics and Kavli Institute for Theoretical Physics China (KITPC),
Institute of Theoretical Physics, Chinese Academy of Sciences, Beijing 100190, P. R. China}

\affiliation{ School of Physical Sciences, University of Chinese Academy of Sciences,
  Beijing 100049, P. R. China}

\affiliation{School of Physical Electronics, University of Electronic Science and Technology of China,
Chengdu 610054, P. R. China}

\begin{abstract}

The natural supersymmetry (SUSY) requires light stop quarks, light sbottom quark,
and gluino to be around one TeV or lighter. The first generation squarks can be effectively large which does not introduce any hierarchy problem in order to escape the constraints from LHC. In this paper we consider a Yukawa deflect medation to realize the effective natural supersymmetry where the interaction between squarks and messenger are made natural under certain Frogget-Nelson $U(1)_X$ charge. The first generation squarks obtain large and postive contribution from the yukawa deflect mediation. The corresponding phenomenology as well as sparticle spectrum are discussed in detail.

\end{abstract}

\maketitle

\section{Introduction}
\label{sec:intro}
Gauge Mediated SUSY Breaking (GMSB)~\cite{Giudice:1998bp} is an elegant framework. In its minimal form, the SUSY breaking hidden sector can be communicated with visible sector only through usual gauge interaction. Which can be realized by introducing spurion field $X$ with $\langle X \rangle=M+\theta^2 F$ and messenger fields $\Phi$. Corresponding superpotential is written as
\begin{align}
W&=X\Phi\bar\Phi\;.
\end{align}
Here spurion $X$ couples to the SUSY breaking sector and $\langle X \rangle$ parameterizes the SUSY breaking effects, $\Phi$ are charged under the Standard Model (SM) $SU(3)\times SU(2)\times U(1)$ gauge group. Since the mass matrix of scalar messenger components are not supersymmetric, the SUSY breaking effects from hidden sector can be mediated to visible sectors via messenger loops.
Compared with gravity mediated SUSY breaking, GMSB has two obvious advantages:
\begin{itemize}
  \item Soft terms are fully calculable. Even in the case of strongly coupled hidden sector, the soft terms can be still expressed as simple correlation functions of hidden sector, namely the scenario of General Gauge Mediation (GGM)~\cite{Meade:2008wd}.
  \item It is inherently flavor-conserving since gauge interaction is flavor-blinded, thus is strongly motivated by the SUSY flavor problem.
\end{itemize}

However, the status of minimal GMSB has been challenged after the discovery of SM-like higgs boson with mass of $125$ GeV~\cite{Draper:2011aa,Ajaib:2012vc}. In order to lift higgs mass to such desirable range, it then implies that higgs mass should be received significant enhancement either from radiative corrections via stop/top loops~\cite{Heinemeyer:2011aa,Arbey:2011ab} or from extra tree-level sources~\cite{Hall:2011aa}. The first option can be achieved through extremely heavy and unmixed stops, or through lighter stops with maximal mixing (large trilinear soft term of stops)~\cite{Casas:1994us,Carena:1995bx,Haber:1996fp}. While in minimal GMSB, the vanishing trilinear soft term at the messenger scale leads to maximal mixing is impossible. The second option requires the extension of Minimal Supersymmetric Standard Model (MSSM) and has been widely investigated~\cite{Ellwanger:2011aa,FileviezPerez:2012iw,Boudjema:2012cq,Jeong:2012ma,
An:2012vp,Hardy:2012ef,Hirsch:2012kv,Basak:2012bd,SchmidtHoberg:2012yy,Benakli:2012cy,
Lu:2013cta,Bertuzzo:2014bwa,Staub:2014pca,Ding:2015wma,Alvarado:2015yna,Bandyopadhyay:2015oga}. In this paper, we consider the first option where large trilinear term is required to soften fine-tuning. In fact, if the messenger sector is allowed to couple with squark or higgs, the problem is improved with trilinear soft terms generated by the additional interaction. This type of interactions relevant to generate large trilinear terms can be generally divided into two types, i.e., higgs mediation and squark mediation. However, higgs mediation generates irreducible positive contribution $\delta m_{H_u}^2\sim A^2_{H_u}$ and leads to large fine-tuning, which is the so-called $A/m^2_{H_u}$ problem. The situation is quite different in squark mediation since it does not suffer from such problem thus has better control on fine-tuning. As a price, squark mediation reintroduces dangerous flavor problem since there is no prior reason to specify the hierarchy and alignment of yukawa matrix of squark. In this direct, Froggatt-Nielsen (FN) mechanism~\cite{Froggatt:1978nt} is adopted as a canonical solution. Here we take the same strategy for squark mediation. In a previous study, Ref.~\cite{Shadmi:2011hs} considered the type of sfermion-sfermion-messenger interaction with FN mechanism. In this work, we extend the model to include sfermion-messenger-messenger interaction and exam its
phenomenology systematically.

The rest of this paper is layout as follows. In section~\ref{sec:model}, we present our notation and model contents. The realization of FN mechanism in Supersymmetric Standard Models (SSMs) and $SU(5)$ models is reviewed in section~\ref{sec:FN}. In section~\ref{sec:mediation}, The FN mechanism is extended to constrain the possible interactions between squarks and messengers. We show that a unique interaction can be obtained with appropriate charge assignment. In section~\ref{sec:pheno}, we explore the phenomenology of this model with emphasize on spectra and fine-tuning issues. The last section is devoted to conclusion and discussion.

\section{Vector-Like Particles (Messengers) in the SSMs and $SU(5)$ Models}
\label{sec:model}
First, we list our convention for SSMs. We denote the left-handed quark doublets, right-handed up-type quarks, right-handed  down-type quarks, left-handed lepton doublets, right-handed neutrinos, and right-handed charged leptons as $Q_i$, $U^c_i$, $D^c_i$, $L_i$, $N^c_i$, and  $E^c_i$, respectively. Also, we denote one pair of higgs doublets as $H_u$ and $H_d$, which give masses to the up-type quarks/neutrinos and the down-type quark/charged leptons, respectively.

In this paper, we consider the messenger particles as the
vector-like particles whose quantum numbers are the same as those of the SM fermions and their Hermitian conjugates. As we know, the generic vector-like particles do not need to form complete $SU(5)$ or $SO(10)$ representations  in Grand Unified Theories (GUTs) from the orbifold constructions~\cite{kawa, GAFF, LHYN, AHJMR, Li:2001qs, Dermisek:2001hp, Li:2001tx, Gogoladze:2003ci},
intersecting D-brane model building on Type II orientifolds~\cite{Blumenhagen:2005mu, Cvetic:2002pj, Chen:2006ip},
M-theory on $S^1/Z_2$ with Calabi-Yau
compactifications~\cite{Braun:2005ux, Bouchard:2005ag}, and
F-theory with $U(1)$ fluxes~\cite{Vafa:1996xn,
Donagi:2008ca, Beasley:2008dc, Beasley:2008kw, Donagi:2008kj,
Font:2008id, Jiang:2009zza, Blumenhagen:2008aw, Jiang:2009za,
Li:2009cy} (For details, see Ref.~\cite{Li:2010hi}).
Therefore, we will consider two kinds of supersymmetric models:
(1) The SSMs with vector-like particles whose $U(1)_X$ charges can be
completely different; (2) The $SU(5)$ Models.

In the SSMs, we introduce the following vector-like particles whose
quantum numbers under $SU(3)_C \times SU(2)_L \times U(1)_Y$ are given explicitly as follows
\begin{eqnarray}
&& XQ + XQ^c = {\mathbf{(3, 2, {1\over 6}) + ({\bar 3}, 2,
-{1\over 6})}}\,;\\
&& XU + XU^c = {\mathbf{ ({3},
1, {2\over 3}) + ({\bar 3},  1, -{2\over 3})}}\,;\\
&& XD + XD^c = {\mathbf{ ({3},
1, -{1\over 3}) + ({\bar 3},  1, {1\over 3})}}\,;\\
&& XL + XL^c = {\mathbf{(1,  2, {-1\over 2}) + ({1},  2,
{1\over 2})}}\,;\\
&& XE + XE^c = {\mathbf{({1},  1, {-1}) + ({1},  1,
{1})}}\,.
\end{eqnarray}

In the $SU(5)$ models, we have three families of the SM fermions
whose quantum numbers under $SU(5)$ are
\bea
F_i=\mathbf{10}\,,~ {\overline f}_i={\mathbf{\bar 5}}\,,~
\label{SU(5)-smfermions}
\eea
where $i=1, 2, 3$ for three families.
The SM particle assignments in $F_i$ and ${\bar f}_i$  are
\bea
F_i=(Q_i, U^c_i, E^c_i)~,~{\overline f}_i=(D^c_i, L_i)~.~
\label{SU(5)-smparticles}
\eea
To break the $SU(5)$ gauge symmetry and electroweak gauge symmetry,
we introduce the adjoint Higgs field and one pair
of Higgs fields whose quantum numbers under $SU(5)$ are
\bea
\Phi~=~ {\mathbf{24}}~,~~~
H~=~{\mathbf{5}}~,~~~{\overline H}~=~{\mathbf{\bar {5}}}~,~\,
\label{SU(5)-1-Higgse}
\eea
where $H$ and ${\overline H}$ contain the Higgs doublets
 $H_u$ and $H_d$, respectively.

We consider the vector-like particles which form complete
 $SU(5)$ multiplets.
The quantum numbers for these additional vector-like particles
 under the $SU(5)\times U(1)_X$ gauge symmetry are
\begin{align}
XF ~=~{\mathbf{10}}~,~~{\overline{XF}}~=~{\mathbf{{\overline{10}}}}~,~
Xf~=~{\mathbf{5}}~,~~{\overline{Xf}}~=~{\mathbf{{\overline{5}}}}~.
\end{align}
The particle contents for the decompositions
of $XF$, ${\overline{XF}}$, $Xf$, and ${\overline{Xf}}$ under the SM
gauge symmetries are
\begin{eqnarray}
&& XF ~=~ (XQ, XU^c, XE^c)~,~~ {\overline{XF}}~=~(XQ^c, XU, XE)~,~\\
&& Xf~=~(XD, XH_u)~,~ {\overline{Xf}}~=~ (XD^c, XH_d)~.~\,
\end{eqnarray}
When we introduce two pairs of  $Xf$ and ${\overline{Xf}}$,
we denote them as $Xf_i$ and ${\overline{Xf}}_i$ with $i=1,~2$.

In this paper, we consider the messenger parity, for example, discrete
$Z_n$ symmetry with $n \geq 2$. Under this $Z_n$ symmetry, the vector-like
particles $X\Phi$ and $X\Phi^c$ transform as follows
\begin{eqnarray}
X\Phi \rightarrow \omega X\Phi~,~~X\Phi^c \rightarrow \omega^{n-1} X\Phi^c~,~\,
\end{eqnarray}
where $\omega^n=1$. Thus, the lightest messenger will be stable.
If the reheating temperature is lower than the mass of the
lightest messenger, there is no cosmological problem. This is indeed
work in our models. Otherwise,
we can break the messenger parity a little bit by turning on tiny VEVs for
$XL$ and/or $XL^c$.

In the gauge mediation, it is very difficult to obtain the Higgs boson with
mass around 125.5~GeV due to the small top quark trilinear soft $A_t$ term
unless the stop quarks are very heavy around 10 TeV.
To generate the large top quark trilinear soft $A_t$ term, we introduce the
superpotential term $XQ XU^c H_u$~\cite{Kang:2012ra, Craig:2012xp}.
In addition, we consider high scale gauge mediation by choosing
\begin{eqnarray}
\langle S \rangle \sim 10^{14}~{\rm GeV}~,~~ F_S \sim 10^{20} ~{\rm GeV}~.~
\end{eqnarray}
The point is that we can increase the magnitude of top quark trilinear soft term
via RGE running. Another point is that the couplings between the spurion and
messengers can be very small because $F_S/\langle S \rangle^2 \sim 10^{-8}$.

\section{Froggatt-Nielsen Mechanism via An Anomalous $U(1)_X$ Gauge Symmetry}
\label{sec:FN}

It is well known that the SM fermion masses and mixings can be explained elegantly via the FN mechanism, where an additional flavor dependent global $U(1)_X$ symmetry is introduced. To stabilize this mechanism against
quantum gravity corrections, we consider an anomalous gauged $U(1)_X$ symmetry.
In a weakly coupled heterotic string theory, there exists an
anomalous $U(1)_X$ gauge symmetry where the corresponding
anomalies are cancelled by the Green-Schwarz mechanism~\cite{MGJS}.
For simplicity, we will not consider the $U(1)_X$ anomaly cancellation
here, which can be done in general by introducing extra vector-like particles
as in Refs.~\cite{Dreiner:2003hw,Dreiner:2003yr,Dreiner:2006xw, Gogoladze:2007ck}.

To break the $U(1)_X$ gauge symmetry, we introduce a flavon field
$A$ with $U(1)_X$ charge $-1$. To preserve SUSY close to the string scale, $A$ can acquire a VEV so that the $U(1)_X$ D-flatness can
be realized.  It was shown~\cite{Dreiner:2003hw, Dreiner:2003yr} that
\begin{equation}
0.171 \le \epsilon\equiv{\frac{\langle A \rangle}{M_{\rm Pl}}} \le 0.221
~,~\,
\end{equation}
where $M_{\rm Pl}$ is the reduced Planck scale.
Interestingly, $\epsilon$ is about the size of the Cabibbo angle.
Also, the $U(1)_X$ charges of the SM fermions and the
Higgs fields $\phi$ are denoted as $Q^X_{\phi}$.

The SM fermion Yukawa coupling terms arising from
the holomorphic superpotential at the string scale in the SSMs are given by
\begin{eqnarray}
- \mathcal{L} &=& y^U_{ij} \left({\frac{A}{M_{\rm Pl}}}\right)^{XYU_{ij}}
Q_i U^c_j H_u+
y^D_{ij} \left({\frac{A}{M_{\rm Pl}}}\right)^{XYD_{ij}} Q_i D^c_j H_d
 \nonumber\\ &&
+y^E_{ij} \left({\frac{A}{M_{\rm Pl}}}\right)^{XYE_{ij}} L_i E^c_j H_d+
y^{N}_{ij} \left({\frac{A}{M_{\rm Pl}}}\right)^{XYN_{ij}} L_i N^c_j
H_u~,~\,
\end{eqnarray}
where $y^U_{ij}$, $y^D_{ij}$, $y^E_{ij}$, and $y^{N}_{ij}$ are
order one Yukawa couplings, and $XYU_{ij}$, $XYD_{ij}$, $XYE_{ij}$
and $XYN_{ij}$ are non-negative integers:
\begin{eqnarray}
&& XYU_{ij} = Q^X_{Q_i} + Q^X_{U^c_j} + Q^X_{H_u}~,~~
XYD_{ij} = Q^X_{Q_i} + Q^X_{D^c_j} + Q^X_{H_d}~,~\nonumber\\
&& XYE_{ij} = Q^X_{L_i} + Q^X_{E^c_j} + Q^X_{H_d}~,~
XYN_{ij} = Q^X_{L_i} + Q^X_{N^c_j} + Q^X_{H_u}~.~\,
\end{eqnarray}
Similarly, the  SM fermion Yukawa coupling terms in the $SU(5)$ models are
\begin{eqnarray}
- \mathcal{L} &=& y^U_{ij} \left({\frac{A}{M_{\rm Pl}}}\right)^{XYU_{ij}}
F_i F_j H +
y^{DE}_{ij} \left({\frac{A}{M_{\rm Pl}}}\right)^{XYDE_{ij}} F_i {\bar f}_j \overline{H}
 \nonumber\\ &&
+ y^{N}_{ij} \left({\frac{A}{M_{\rm Pl}}}\right)^{XYN_{ij}}  {\bar f}_i N^c_j H~,~\,
\end{eqnarray}
where
\begin{eqnarray}
&& XYU_{ij} = Q^X_{F_i} + Q^X_{F_j} + Q^X_{H}~,~~
XYDE_{ij} = Q^X_{F_i} + Q^X_{{\bar f}_j} + Q^X_{\overline{H}}~,~\nonumber\\
&& XYN_{ij} = Q^X_{{\bar f}_i} + Q^X_{N^c_j} + Q^X_{H}~.~\,
\end{eqnarray}

In addition, we shall employ
the quark textures for the SSMs and $SU(5)$ models in Table~\ref{tab:textures},
which can reproduce the SM quark Yukawa couplings and the CKM quark mixing
matrix for $\epsilon \approx 0.2$~\cite{Dreiner:2003hw,Dreiner:2003yr,Dreiner:2006xw}.
\begin{table}
\[
\begin{array}{|c|c|c|}
\hline
\text{Yukawa}   & \text{The SSMs}    & \text{$SU(5)$ Models}  \\
\hline
Y^U     &
\begin{pmatrix}\epsilon^{8}&\epsilon^{5}&\epsilon^{3}\\
\epsilon^{7}&\epsilon^{4}&\epsilon^{2}\\
\epsilon^{5}&\epsilon^{2}&\epsilon^{0}\end{pmatrix}     &
\begin{pmatrix}\epsilon^{6}&\epsilon^{5}&\epsilon^{3}\\
\epsilon^{5}&\epsilon^{4}&\epsilon^{2}\\
\epsilon^{3}&\epsilon^{2}&\epsilon^{0}\end{pmatrix}    \\
\hline
Y^D     &
\epsilon^c\begin{pmatrix}\epsilon^{4}&\epsilon^{3}&\epsilon^{3}\\
\epsilon^{3}&\epsilon^{2}&\epsilon^{2}\\
\epsilon^{1}&\epsilon^{0}&\epsilon^{0}\end{pmatrix}     &
\epsilon^c\begin{pmatrix}\epsilon^{4}&\epsilon^{3}&\epsilon^{3}\\
\epsilon^{3}&\epsilon^{2}&\epsilon^{2}\\
\epsilon^{1}&\epsilon^{0}&\epsilon^{0}\end{pmatrix}  \\
\hline
\end{array}
\]
\caption{The quark textures in the SSMs and $SU(5)$ models.}
\label{tab:textures}
\end{table}
And the following lepton textures can reproduce the neutrino masses and PMNS
neutrino mixing matrix:
\begin{align}
Y^E \sim \epsilon^c\begin{pmatrix}\epsilon^{4}&\epsilon^{3}&\epsilon^{1}\\
\epsilon^{3}&\epsilon^{2}&\epsilon^{0}\\
\epsilon^{3}&\epsilon^{2}&\epsilon^{0}\end{pmatrix}\;,\quad
M_{LL} \sim \frac{\langle H_u \rangle^2}{M_s}
\epsilon^{-5}\begin{pmatrix}\epsilon^{2}&\epsilon^{1}&\epsilon^{1}\\
\epsilon^{1}&\epsilon^{0}&\epsilon^{0}\\
\epsilon^{1}&\epsilon^{0}&\epsilon^{0}\end{pmatrix}\;,
\label{eq:neutrino}
\end{align}
where $c$ is either 0, 1, 2 or 3, and
$\tan \beta \equiv \langle H_u \rangle / \langle
H_d \rangle$ satisfies $\epsilon^c \sim \epsilon^3 \tan \beta$. This neutrino
texture requires some amount of fine-tuning as it generically predicts
\begin{eqnarray}
\sin \theta_{12} \sim \epsilon  ~,~~
\Delta m^2_{12} \sim \Delta m^2_{23}~.~\,
\end{eqnarray}
Interestingly, with $\epsilon$ as large as $0.2$, the
amount of fine-tuning needed is
not that huge and this is shown in the computer simulations of
\cite{Dreiner:2003hw,Dreiner:2003yr,Dreiner:2006xw} with random values for the
coefficients.

To be concrete, we choose the $U(1)_X$ charges for the SM fermions and Higgs fields
in the SSMs as follows
\begin{eqnarray}
&& Q^X_{Q_i} ~=~(3,2,0)~,~~ Q^X_{U^c_i} ~=~(5,2,0)~,~~ Q^X_{D^c_i} ~=~(c+1,~c,~c)~,~\nonumber \\
&& Q^X_{L_i} ~=~(c+1,~c,~c)~,~~Q^X_{E^c_i} ~=~(3,2,0)~,~~Q^X_{H_u}~=~Q^X_{H_d} ~=~0~,~\,
\end{eqnarray}
with $Q^X_{\phi_i} \equiv (Q^X_{\phi_1}, ~Q^X_{\phi_2},~Q^X_{\phi_3}) $ for the SM fermions $\phi_i$.

Also, we take the following $U(1)_X$ charges for the SM fermions and Higgs fields in
the $SU(5)$ models
\begin{align}
Q^X_{F_i} ~=~(3,2,0)~,~~ Q^X_{{\bar f}_i}~=~(c+1,~c,~c)~,~~ Q^X_{H}~=~Q^X_{\overline{H}} ~=~0~.
\end{align}

\section{Squark Mediation vs Higgs Mediation}
\label{sec:mediation}
Natural SUSY can be regarded as an effective SUSY scenario where only stop, gluino and small $\mu$ term are required in the spectra. As a consequence, the fine-tuning remains a manageable level. One of nice property of Natural SUSY is that the first two generation squarks can be very heavy without introducing any fine-tuning, which also evade bounds of SUSY direct searches from LHC. In terms of squark mediation with squark-messsenger-messenger interaction, Squarks receive additional positive contribution thus is possible to construct Natural SUSY model.

The basic formulas to compute corresponding soft terms are given as~\cite{Evans:2013kxa},
\begin{align}
\label{eq:Amsqfinal}
A_{ab} &  =  -{1\over 32 \pi^2}d_{a}^{ij} \Delta \left( \lambda^*_{a ij} \lambda_{b i j } \right)\Lambda\;,\\
\delta m_{ab}^2 &= {1\over256\pi^4}\Bigg(
{1\over2}d_a^{c B}d_{B}^{de} \lambda_{a c B}^*\lambda _{b c C}\lambda_{de B}\lambda_{d e C}^*
+ d_a^{cB}d_c^{d C}\lambda_{a c B}^*\lambda_{b e B}\lambda_{c d C}\lambda_{ d e C}^* \nonumber\\
& + d_{a}^{c B}d_{b}^{d C}\lambda_{a c B}^*\lambda_{c e B}\lambda_{d e C}^*\lambda_{b d C}
-d_a^{c d}d_c^{fB} y_{a c d}^* y_{b d e }\lambda_{c f B}\lambda_{e f B}^* + {1\over2}d_a^{c B}d_c^{ef}y_{c ef}y_{d ef}^* \lambda_{a c B}^*\lambda_{bd B} \nonumber\\
&  + \frac 12 d_a^{cd}d_c^{ef} y_{acd}^* y_{cef}\lambda_{b d B}\lambda_{ef B}^*
+ \frac 12 d_a^{cB}d_B^{ef} \lambda_{a c B}^* \lambda_{ef B} y_{bcd} y_{def}^*
 -2d_{a}^{c B}C_r^{acB} g_r^2 \lambda_{a c B}^*\lambda_{b c B }
\Bigg)\Lambda^2\;,
\end{align}
where $\Lambda = F/M$, and $C_r^{ijk} = c_r^i+c_r^j+c_r^k$ is the sum of the quadratic Casimirs of each field interacting through $\lambda_{ijk}$. In above expressions, we do not include the contributions from usual GMSB (thus is labeled by $\delta m_{ab}^2$) and all of indices are summed over except for $a$ and $b$. Without the FN mechanism, there will be general interaction between $Q_i$,$U_i$ and $D_i$. The squark mediation is not automatically minimal flavor violation like higgs mediation.  The MSSM-MSSM mixing term gives rise to dangerous non-vanishing and non-diagonal soft masses, for example,
\begin{align}
m_{Q_1 Q_2}^2\sim \lambda_{q1}^2\lambda_{q2}^2\Lambda^2\;.
\label{eqn:mixing}
\end{align}
The non-diagonal terms in Eq.~(\ref{eqn:mixing}) motives Ref.~\cite{}{\color{red}}to construct chiral flavor violation scenario where only single $Q_i$ or $U_i$, $D_i$ is allowed to couple the messenger. As a result, the dangerous flavor violation term is suppressed naturally. However our situation does not belong to chiral flavor violation. In order to realize effective SUSY scenario, all the first and second generation squarks must be coupled to messengers in order to obtain large soft masses enhancement. It seems the non-diagonal term is inevitable in Eq.~(\ref{eqn:mixing}).  The loop hole comes from the fact that the bound is greatly improved once the squark are not degenerate. In particular, the largest bound comes from the first generation squarks because of large PDF effct of first generation quarks. Therefore it strongly suggests us to consider the first generation squark mediation which is technically natural under FN mechanism. That is the basic motivtion for us to consider FN mechanism in squark mediation. The FN natural model is free from MSSM-MSSM mixing and the formula is reduced to
\begin{align}
A_{a}& = -{1\over16\pi^2}d_{a}^{c B} \lambda_{acB}^2\Lambda\;,\nonumber\\
\delta m_{a}^2 &= {1\over256\pi^4}\Bigg(
{1\over2}d_a^{ c B }d_{B}^{de} |\lambda_{a c B}|^2 |\lambda_{de B}|^2
+ d_a^{c B}d_c^{dC}|\lambda_{acB}|^2 |\lambda_{c dC}|^2 \nonumber\\
& + d_{a}^{cB}d_{a}^{dC}|\lambda_{acB}|^2 |\lambda_{adC}|^2
  -d_a^{cd}d_c^{f B} | y_{acd}|^2 |\lambda_{c f B}|^2 + {1\over2}d_a^{c B}d_c^{ef} |y_{cef}|^2 |\lambda_{acB}|^2\nonumber\\
& + \frac 12  d_a^{cd}d_c^{ef} y_{acd}^* y_{cef}\lambda_{a dB}\lambda_{efB}^*  + \frac 12 d_a^{cB}d_{B}^{ef} \lambda_{ac B}^*\lambda_{efB}y_{acd} y_{d ef}^*
 -2d_{a}^{cB}C_r^{acB}g_r^2  |\lambda_{a cB}|^2
\Bigg)\Lambda^2\;.
\label{eq:typeIIm}
\end{align}

Let us demonstrate how FN mechanism makes the squark mediation flavor blinded. The general  squark-messenger-messenger interaction within the messenger sector being $SU(5)$ complete multiplets is divided into $Q$-type, $U$-type and $D$-type mediations, here $U$ and $D$ respectively denote $\bar u$ and $\bar d$ for short. In table~\ref{tab:charge}, we list the complete messenger fields and their $U(1)$ charge assignment.
\begin{table*}[hbtp]
\begin{tabular}{|c|c|c|c|c|c|c|}
\hline
Messenger & $(\text{XQ},~\text{XQ}^c)$ & $(\text{XU},~\text{XU}^c)$& $(\text{XL},~\text{XL}^c)$& $(\text{XD},~\text{XD}^c)$ & $(\text{XE},~\text{XE}^c)$& $\text{XS}$\\\hline
$U(1)$ Charge & $(3,~-3)$ & $(-5,~5)$ & $(2,~-2)$
   & $(3,~-3)$ & $(0,~0)$ & $0$\\\hline
\end{tabular}
\caption{Complete list of messenger fields and their $U(1)$ charge assignment.}
\label{tab:charge}
\end{table*}

For the Q-type Mediation, the most general superpotential is
\begin{align}
W_{Q}=\lambda_{q1_i} Q_{i} XQ^{c} XS +\lambda_{q2_i} Q_{i} XD^c XL + \lambda_{q3_i} Q_{i} XU^c XL^c+ \lambda_{q4_i} XQ XD\;,
\end{align}
where $i=1,...,3$ is family indices. Based on table~\ref{tab:charge}, the Yukawa couplings in Q-type mediation can be determined as below,
\begin{align}
\lambda_{q1_i}&\sim \left\{1,\frac{1}{\epsilon
   },\frac{1}{\epsilon ^3}\right\}\;,\quad\lambda_{q2_i}\sim \left\{\epsilon ^2,\epsilon
   ,\frac{1}{\epsilon }\right\}\;,\nonumber\\
\lambda_{q3_i}&\sim\left\{\epsilon ^6,\epsilon ^5,\epsilon^3\right\}\;,
\quad\lambda_{q4_i}\sim \left\{\epsilon ^9,\epsilon ^8,\epsilon^6\right\}\;.
\end{align}
Terms with negative order of $\epsilon$ must be removed in order not to violate the holomorphy requirement of superpotential. While terms with positive order of $\epsilon$ can be ignored which is guaranteed by the smallness of $\epsilon$. Therefore Only $\lambda_{q1_1}$ is allowed under the consideration of FN mechanism and holomorphy.  For now we only consider squark-messenger-messenger interaction, this is mainly because the squark-squark-messenger interaction under FN charges has been discussed in the literature. Since only the $Q_1$ mediation is allowed, there is no flavor-changing problem. For the $U$-type mediation the most general superpotential is
\begin{align}
W_U=\lambda_{u1_i} U_i XU XS+\lambda_{u2_i} U_i XD^c XD^c+ \lambda_{u3_i} U XQ XL^c+\lambda_{u4_i} U_i XE XD\;.
\end{align}

According to FN mechanism the coupling scales like
\begin{align}
\lambda_{u1_i}&\sim \left\{1,\frac{1}{\epsilon
   ^3},\frac{1}{\epsilon ^5}\right\}\;,\quad
\lambda_{u2_i}\sim \left\{\frac{1}{\epsilon
   },\frac{1}{\epsilon
   ^4},\frac{1}{\epsilon ^6}\right\}\;,\nonumber\\
\lambda_{u3_i}&\sim \left\{\epsilon ^6,\epsilon ^3,\epsilon
   \right\}\;,\quad
\lambda_{u4_i}\sim \left\{\epsilon ^2,\frac{1}{\epsilon
   },\frac{1}{\epsilon ^3}\right\}\;.
\end{align}
It is similar with Q-type mediation, only the $\lambda_{u1_1}$ is allowed. For D-type mediation we have
\begin{align}
W_D=\lambda_{d1_i} D_i XQ XL^c+\lambda_{d2_i} D_i XQ^c XQ^c + \lambda_{d3_i} D_i XD^c XU^c+\lambda_{d4_i} D_i XE^c XU\;.
\end{align}
Subject to the FN mechanism we obtain the couplings
\begin{align}
\lambda_{d1_i}&\sim \left\{\epsilon ^6,\epsilon ^5,\epsilon
   ^5\right\}\;,\quad
\lambda_{d2_i}\sim \left\{\frac{1}{\epsilon
   },\frac{1}{\epsilon
   ^2},\frac{1}{\epsilon ^2}\right\}\;,\nonumber\\
\lambda_{d3_i}&\sim \left\{\epsilon ^7,\epsilon ^6,\epsilon
   ^6\right\}\;,\quad
\lambda_{d4_i}\sim \left\{1,\frac{1}{\epsilon
   },\frac{1}{\epsilon }\right\}\;.
\end{align}

All in all the allowed yukawa defelect mediation interaction for squarks are summarized as follows
\begin{align}
W=\lambda_{q} Q_{1} XQ^{c} XS+\lambda_{u} U_i XU XS +\lambda_{d} D_i XE^c XU\;.
\label{eqn:squark}
\end{align}
From Eq.~(\ref{eqn:squark}), we obtain the extra contribution to soft masses for the first generation squarks. In other words there is no desirable large trilinear term $A_t$ from equation~\ref{eqn:squark} which motivates us to resort to higgs Mediation. Based on FN mechanism the only allowed superpotential for Higgs mediation is
\begin{align}
W_H=\lambda_h H_u XD^c XQ\;.
\label{eqn:higgs}
\end{align}
It is automatically preserves minimal flavor violation (MFV). Using Eq.~(\ref{eq:typeIIm}), we obtain following soft terms
\begin{align}
A_t&=-\frac{3 \Lambda  \lambda _h^2}{16 \pi
   ^2}\nonumber\\
\delta m_{H_u}^2&=\frac{\Lambda ^2 \left(18 \lambda _h^4-6
   \left(\frac{7 g_1^2}{30}+\frac{3
   g_2^2}{2}+\frac{8 g_3^2}{3}\right)
   \lambda _h^2\right)}{256 \pi ^4}
\;,\nonumber\\
\delta m_{Q_3}^2&=-\frac{3 \Lambda ^2 \lambda _h^2
   y_t^2}{256 \pi ^4}\;,\nonumber\\
\delta m_{U_3}^2&=-\frac{3 \Lambda ^2 \lambda _h^2
   y_t^2}{128 \pi ^4}\;,\nonumber\\
\delta m_{Q_1}^2&=\frac{\Lambda ^2 \left(8 \lambda _q^4-2
   \left(\frac{g_1^2}{30}+\frac{3
   g_2^2}{2}+\frac{8 g_3^2}{3}\right)
   \lambda _q^2\right)}{256 \pi ^4}\;,\nonumber\\
\delta m_{U_1}^2&=\frac{\Lambda ^2 \left(5 \lambda _u^4-2
   \left(\frac{13 g_1^2}{30}+\frac{3
   g_2^2}{2}+\frac{8 g_3^2}{3}\right)
   \lambda _u^2\right)}{256 \pi ^4}\;,\nonumber\\
\delta m_{D_1}^2&=\frac{\Lambda ^2 \left(5 \lambda _d^4-2
   \left(\frac{14 g_1^2}{15}+\frac{8
   g_3^2}{3}\right) \lambda
   _d^2\right)}{256 \pi ^4}\;.
\label{eq:soft}
\end{align}
The choice of higgs mediation in Eq.~(\ref{eqn:higgs}) plays a crucial role in reducing the fine-tuning:
\begin{itemize}
\item The trilinear soft term has a overall factor $3$ coming from the higher representation of $SU(5)$. Thus it can give rise to large trilinear term compared with other higgs mediation.
\item The $m_{H_u}^2$ has a negative contribution from $SU(3)$ gauge coupling. Such a large coupling can reduce the fine-tuning easily.
\end{itemize}
The parameter space is thus determined by the following parameters
\begin{align}
\{\Lambda,\, M,\, \lambda_q,\,\lambda_u,\,\lambda_d,\,\lambda_h, \tan\beta, sign(\mu)\}
\label{eqn:input}
\end{align}

\section{Phenomenology Analysis}
\label{sec:pheno}
In this section, we give a detailed discussion on the numerical results for our effective supersymmetry model. In particular the higgs mass, stop mass, gluino mass as well as fine-tuning are given explicitly. In our numerical analysis, the relevant soft terms are firstly generated at messenger scale in terms of gauge  mediation and higgs, squark mediation. The low scale soft terms are obtained by solving the two-loop RG equations. For this purpose, we implemented the corresponding boundary conditions in Eq.~(\ref{eq:soft}) into the Mathematica package {\tt SARAH}~\cite{Staub:2008uz,Staub:2009bi,Staub:2010jh,Staub:2012pb,Staub:2013tta}. Then {\tt SARAH} is used to create a {\tt SPheno}~\cite{Porod:2003um,Porod:2011nf} version for the MSSM to calculate particle spectrum. The tasks of parameter scans are implemented by package~{\tt SSP}~\cite{Staub:2011dp}.

The framework that we concentrate on is MSSM with yukawa deflect mediation. Its input parameters are given in Eq.~(\ref{eqn:input}). The scan range we adapt is
\begin{eqnarray}
\Lambda&\in&(6\times 10^4,~6\times 10^5)~\text{GeV}\;,\nonumber \\
\lambda_h&\in&(0,~1.2)\;.
\end{eqnarray}
The other parameters have been fixed to $M=10^8$ GeV, $\tan\beta=10$ and $sign(\mu)=1$. For the parameters in squark mediation, we divide it into two scenarios:
\begin{eqnarray}
&&\text{Degenerated squark:}\;\lambda_q=\lambda_u=\lambda_d=0\;,\nonumber \\
&&\text{Non-degenerated squark:}\;\lambda_q=\lambda_u=\lambda_d=1.2\;.
\end{eqnarray}
During the scan, various mass spectrum and low energy constraints have been considered and listed at below:
\begin{enumerate}
  \item {The higgs mass constraints:
\begin{align}
 123~{\rm GeV}\leq m_h \leq 127~{\rm GeV}\;,
\end{align}}
  \item {LEP bounds and B physics constraints:
\begin{eqnarray}\label{eqn:Bphysics}
1.6\times 10^{-9} \leq{\rm BR}(B_s \rightarrow \mu^+ \mu^-)
  \leq 4.2 \times10^{-9}\;(2\sigma)~\text{\cite{CMS:2014xfa}}\;,
\nonumber\\
2.99 \times 10^{-4} \leq
  {\rm BR}(b \rightarrow s \gamma)
  \leq 3.87 \times 10^{-4}\;(2\sigma)~\text{\cite{Amhis:2014hma}}\;,
\nonumber\\
7.0\times 10^{-5} \leq {\rm BR}(B_u\rightarrow\tau \nu_{\tau})
        \leq 1.5 \times 10^{-4}\;(2\sigma)~\text{\cite{Amhis:2014hma}}\;.
\end{eqnarray}}
  \item {Sparticle bounds from LHC Run-II:
\begin{itemize}
  \item Light stop mass $m_{{\tilde t}_1} > 850$ GeV~\cite{ATLAS:2016jaa,CMS:2016inz},
  \item Light sbottom mass $m_{{\tilde b}_1}>840-1000$  GeV~\cite{Aaboud:2016nwl,CMS:2016xva},
  \item Degenerated first two generation squarks (both left-handed and right-handed) $m_{{\tilde q}}>1000-1400$ GeV~\cite{CMS:2016xva},
\item Gluino mass $m_{\tilde g} > 1800$ GeV~\cite{ATLAS:2016kts,CMS:2016inz}.
\end{itemize}}
\end{enumerate}

Finally, Barbieri-Giudice measure~\cite{Barbieri:1987fn,Ellis:1986yg} is used to quantify the fine-tuning:
\begin{align}
\label{eq:FT}
\Delta_{\rm FT} \equiv \max \{ \Delta_a\} \;\; \mbox{ where }\;\; \Delta_a \equiv \frac{\partial \log m_Z^2}{\partial \log a} \, .
\end{align}
where $a$ denotes the input parameters in Eq.~(\ref{eqn:input}).

\begin{figure}[!htbp]
\begin{center}
\includegraphics[width=0.6\linewidth]{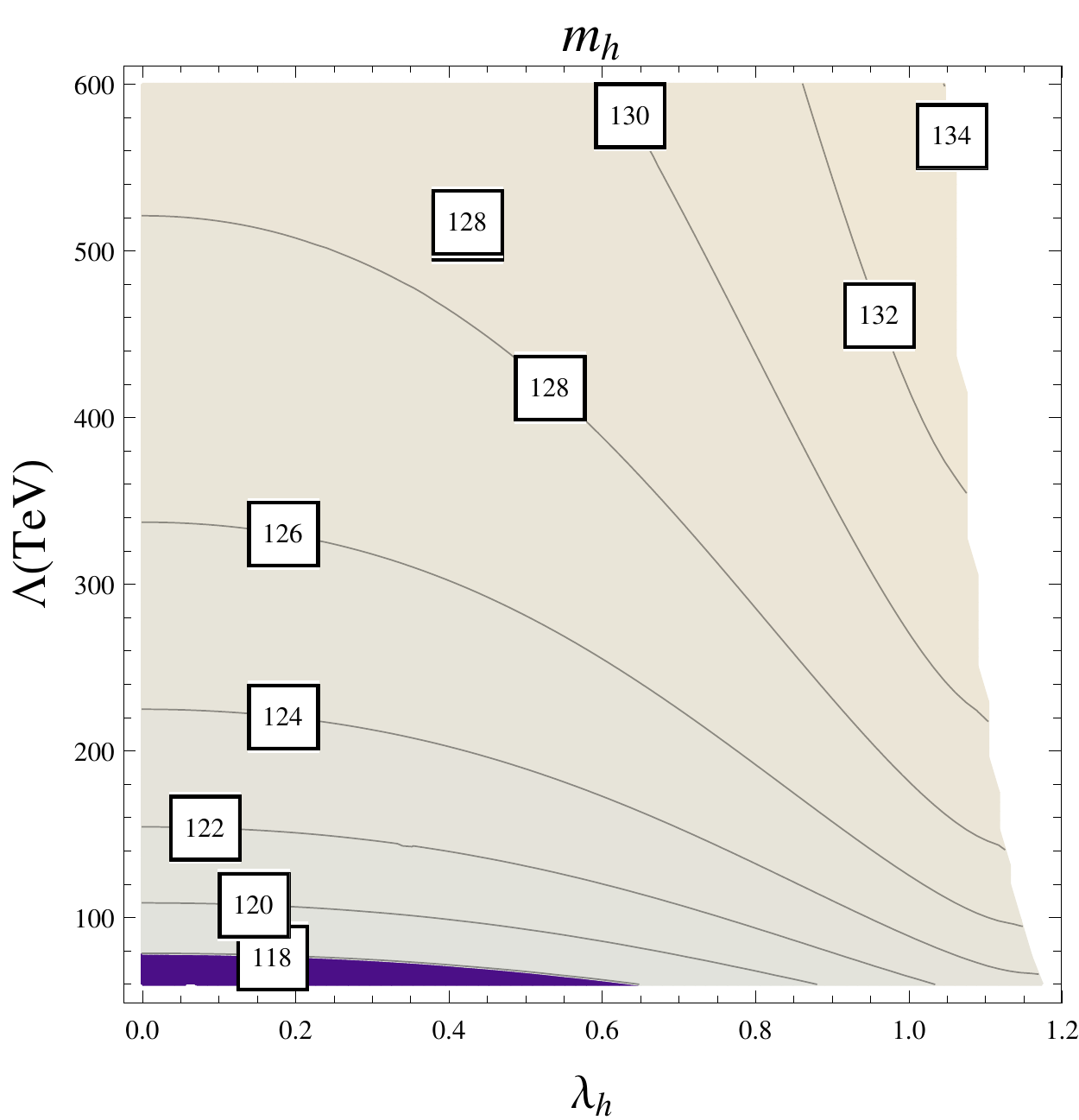}
\end{center}
\caption{Distribution of higgs mass in $[\lambda_h,\Lambda]$ plane.}
\label{fig:higgs}
\end{figure}

\begin{figure}[!htbp]
\begin{center}
\includegraphics[width=0.45\linewidth]{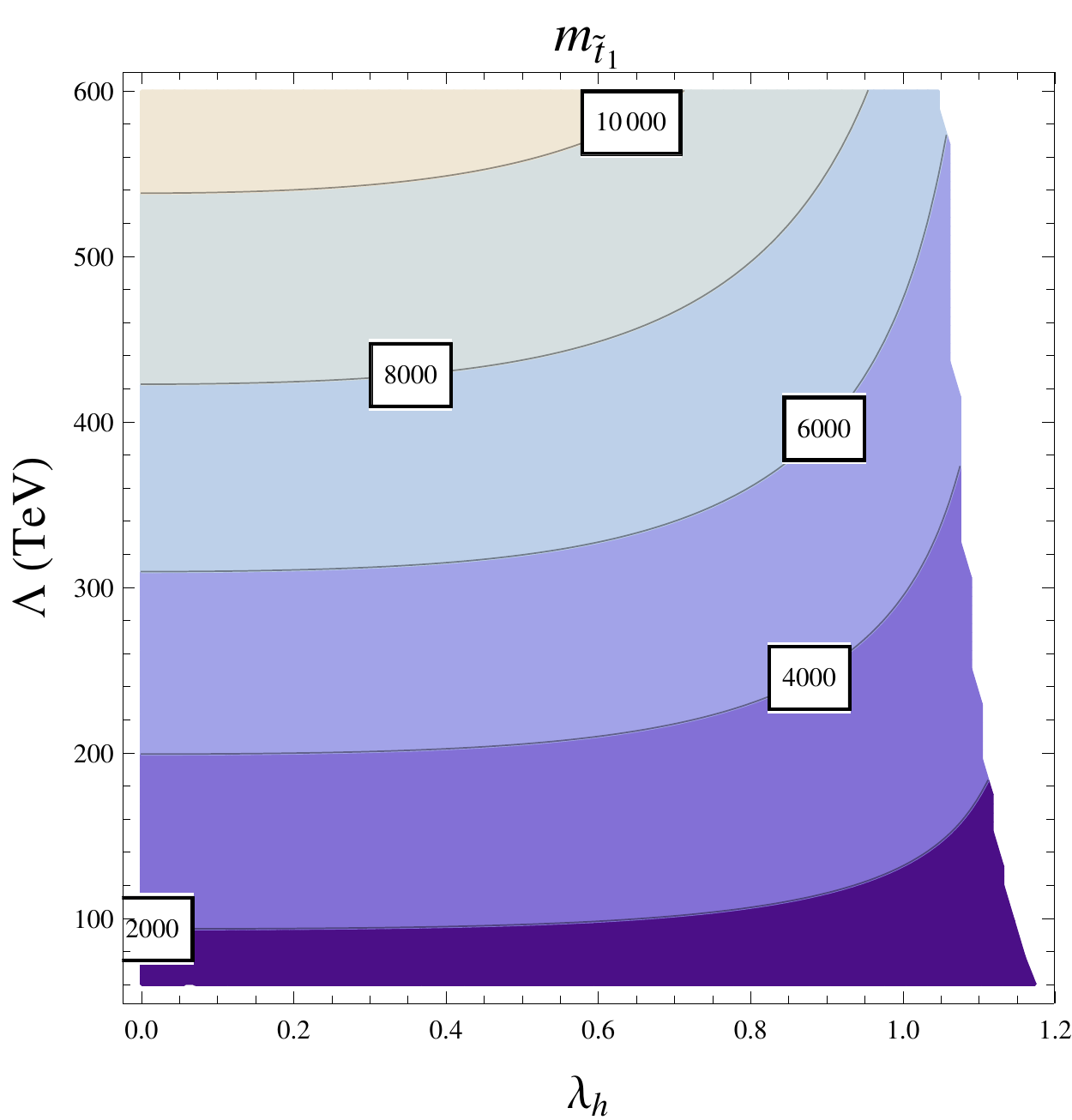}
\includegraphics[width=0.45\linewidth]{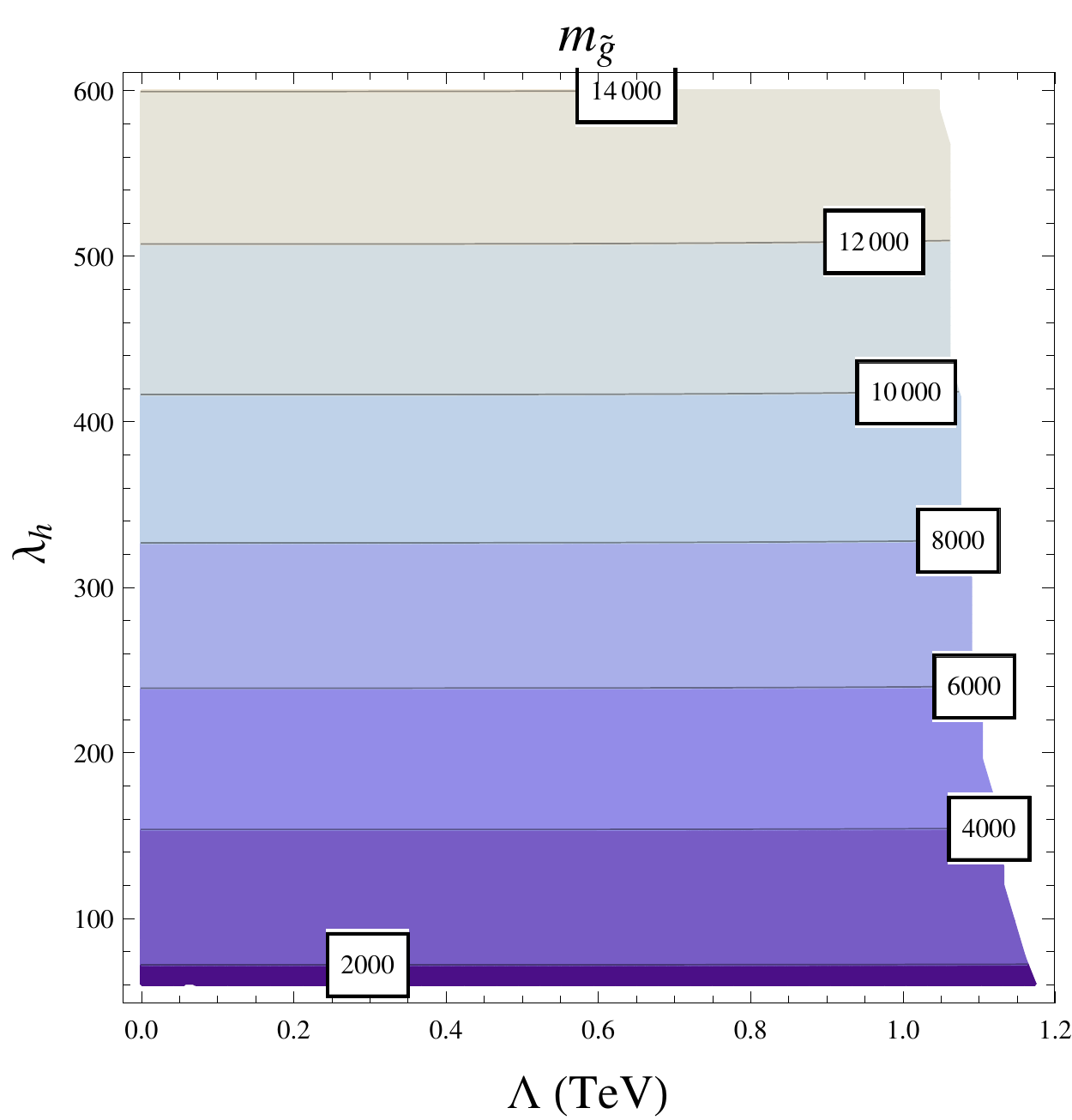}
\end{center}
\caption{Distributions of stop (left-panel) and gluino mass (right-panel) in $[\lambda_h,\Lambda]$ plane.}
\label{fig:stop}
\end{figure}

\begin{figure}[!htbp]
\begin{center}
\includegraphics[width=0.6\linewidth]{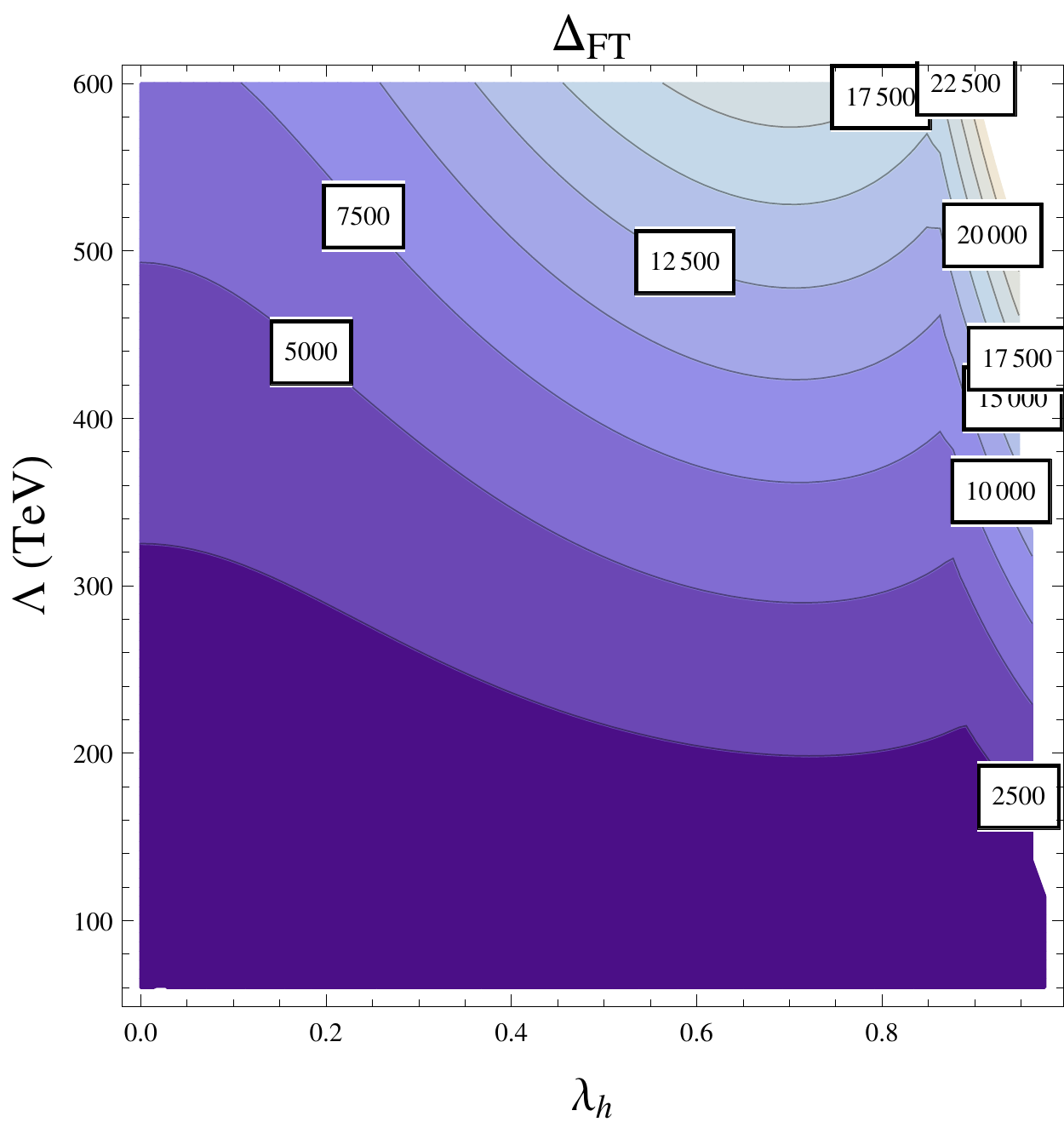}
\end{center}
\caption{Distribution of fine tuning measure in $[\lambda_h,\Lambda]$ plane.}
\label{fig:finetuning}
\end{figure}

In Fig.~\ref{fig:higgs}-\ref{fig:finetuning}, We display  the contour plots of important mass spectra and fine tuning measure $\Delta_{\rm FT}$ in the $[\lambda_{h},\Lambda]$ plane. There are some notable features can be learned from these figures and summarized at below:
\begin{enumerate}
\item {\textbf{The higgs mass:} The higgs mass range is taken from 123 GeV to 127 GeV in our numerical analysis. For small $\lambda_h$, One expects higgs mass simply growth with an increases of $\Lambda$. With the increasing of $\lambda_h$, allowed parameter space is forced to shift to smaller $\Lambda$ region in order to obtain correct higgs mass.}
\item {\textbf{The fine tuning measure:} For small values of $\Lambda$ and $\lambda_h$, $\Delta_{\rm FT}$ is usually dominated by $\Lambda$. Since in these regions the RGE effects are most important, the contribution to the fine tuning of $\lambda_{h}$, which only affects the boundary conditions, is negligible. The important parameters thus is $\Lambda$ which sets the range of the RGE running. For moderate $\Lambda_{a}$ and $\lambda_h$, the contribution from $\mu$ and $\Lambda$ are almost comparable. Finally, if $\lambda_h$ becomes large it is always the biggest contributor to fine tuning measure independent of the value of $\Lambda$.}
\item {\textbf{The squark and gluino masses:} Both stop and gluino masses fall into multi TeV range and therefore out of current LHC reach.}
\end{enumerate}

\section{Conclusions}
In this paper, we have investigated the extended gauge mediation
models where yukawa interaction between messengers and matter superfields
are made natural under the consideration of F-N $U(1)$ symmetry.
Because of higgs mediation the large A-term is generated naturally which can be
used to enhance the higgs mass efficiently.
Considering the additional quark mediation, it is found that first generation
squarks get large positive contribution thus escaping from dangerous LHC constraints.
We also study the parameter space and phenomenology numerically. The results show that
the model is still promising under the stringent LHC constraint.

\section*{Ackownoledgement:}

This research was supported in part  by the Natural Science
Foundation of China under grant numbers 10821504, 11075194,
11135003, and 11275246.


\begin{thebibliography}{0}
\expandafter\ifx\csname natexlab\endcsname\relax\def\natexlab#1{#1}\fi
\expandafter\ifx\csname bibnamefont\endcsname\relax
  \def\bibnamefont#1{#1}\fi
\expandafter\ifx\csname bibfnamefont\endcsname\relax
  \def\bibfnamefont#1{#1}\fi
\expandafter\ifx\csname citenamefont\endcsname\relax
  \def\citenamefont#1{#1}\fi
\expandafter\ifx\csname url\endcsname\relax
  \def\url#1{\texttt{#1}}\fi
\expandafter\ifx\csname urlprefix\endcsname\relax\def\urlprefix{URL }\fi
\providecommand{\bibinfo}[2]{#2}
\providecommand{\eprint}[2][]{\url{#2}}

\end{thebibliography}


\clearpage
\begin{thebibliography}{99}






\bibitem{Giudice:1998bp}
  G.~F.~Giudice and R.~Rattazzi,
  Phys.\ Rept.\  {\bf 322}, 419 (1999)
  doi:10.1016/S0370-1573(99)00042-3
  [hep-ph/9801271].

\bibitem{Meade:2008wd}
  P.~Meade, N.~Seiberg and D.~Shih,
  Prog.\ Theor.\ Phys.\ Suppl.\  {\bf 177}, 143 (2009)
  doi:10.1143/PTPS.177.143
  [arXiv:0801.3278 [hep-ph]].

\bibitem{Draper:2011aa}
  P.~Draper, P.~Meade, M.~Reece and D.~Shih,
  Phys.\ Rev.\ D {\bf 85}, 095007 (2012)
  doi:10.1103/PhysRevD.85.095007
  [arXiv:1112.3068 [hep-ph]].

\bibitem{Ajaib:2012vc}
  M.~A.~Ajaib, I.~Gogoladze, F.~Nasir and Q.~Shafi,
  Phys.\ Lett.\ B {\bf 713}, 462 (2012)
  doi:10.1016/j.physletb.2012.06.036
  [arXiv:1204.2856 [hep-ph]].

\bibitem{Heinemeyer:2011aa}
  S.~Heinemeyer, O.~Stal and G.~Weiglein,
  Phys.\ Lett.\ B {\bf 710}, 201 (2012)
  doi:10.1016/j.physletb.2012.02.084
  [arXiv:1112.3026 [hep-ph]].

\bibitem{Arbey:2011ab}
  A.~Arbey, M.~Battaglia, A.~Djouadi, F.~Mahmoudi and J.~Quevillon,
  Phys.\ Lett.\ B {\bf 708}, 162 (2012)
  doi:10.1016/j.physletb.2012.01.053
  [arXiv:1112.3028 [hep-ph]].

\bibitem{Hall:2011aa}
  L.~J.~Hall, D.~Pinner and J.~T.~Ruderman,
  JHEP {\bf 1204}, 131 (2012)
  doi:10.1007/JHEP04(2012)131
  [arXiv:1112.2703 [hep-ph]].

\bibitem{Casas:1994us}
  J.~A.~Casas, J.~R.~Espinosa, M.~Quiros and A.~Riotto,
  Nucl.\ Phys.\ B {\bf 436}, 3 (1995)
  Erratum: [Nucl.\ Phys.\ B {\bf 439}, 466 (1995)]
  doi:10.1016/0550-3213(94)00508-C, 10.1016/0550-3213(95)00057-Y
  [hep-ph/9407389].

\bibitem{Carena:1995bx}
  M.~Carena, J.~R.~Espinosa, M.~Quiros and C.~E.~M.~Wagner,
  Phys.\ Lett.\ B {\bf 355}, 209 (1995)
  doi:10.1016/0370-2693(95)00694-G
  [hep-ph/9504316].

\bibitem{Haber:1996fp}
  H.~E.~Haber, R.~Hempfling and A.~H.~Hoang,
  Z.\ Phys.\ C {\bf 75}, 539 (1997)
  doi:10.1007/s002880050498
  [hep-ph/9609331].

\bibitem{Ellwanger:2011aa}
  U.~Ellwanger,
  JHEP {\bf 1203}, 044 (2012)
  doi:10.1007/JHEP03(2012)044
  [arXiv:1112.3548 [hep-ph]].

\bibitem{FileviezPerez:2012iw}
  P.~Fileviez Perez,
  Phys.\ Lett.\ B {\bf 711}, 353 (2012)
  doi:10.1016/j.physletb.2012.04.016
  [arXiv:1201.1501 [hep-ph]].

\bibitem{Boudjema:2012cq}
  F.~Boudjema and G.~Drieu La Rochelle,
  Phys.\ Rev.\ D {\bf 86}, 015018 (2012)
  doi:10.1103/PhysRevD.86.015018
  [arXiv:1203.3141 [hep-ph]].

\bibitem{Jeong:2012ma}
  K.~S.~Jeong, Y.~Shoji and M.~Yamaguchi,
  JHEP {\bf 1209}, 007 (2012)
  doi:10.1007/JHEP09(2012)007
  [arXiv:1205.2486 [hep-ph]].

\bibitem{An:2012vp}
  H.~An, T.~Liu and L.~T.~Wang,
  Phys.\ Rev.\ D {\bf 86}, 075030 (2012)
  doi:10.1103/PhysRevD.86.075030
  [arXiv:1207.2473 [hep-ph]].

\bibitem{Hardy:2012ef}
  E.~Hardy, J.~March-Russell and J.~Unwin,
  JHEP {\bf 1210}, 072 (2012)
  doi:10.1007/JHEP10(2012)072
  [arXiv:1207.1435 [hep-ph]].

\bibitem{Hirsch:2012kv}
  M.~Hirsch, W.~Porod, L.~Reichert and F.~Staub,
  Phys.\ Rev.\ D {\bf 86}, 093018 (2012)
  doi:10.1103/PhysRevD.86.093018
  [arXiv:1206.3516 [hep-ph]].

\bibitem{Basak:2012bd}
  T.~Basak and S.~Mohanty,
  Phys.\ Rev.\ D {\bf 86}, 075031 (2012)
  doi:10.1103/PhysRevD.86.075031
  [arXiv:1204.6592 [hep-ph]].

\bibitem{SchmidtHoberg:2012yy}
  K.~Schmidt-Hoberg and F.~Staub,
  JHEP {\bf 1210}, 195 (2012)
  doi:10.1007/JHEP10(2012)195
  [arXiv:1208.1683 [hep-ph]].

\bibitem{Benakli:2012cy}
  K.~Benakli, M.~D.~Goodsell and F.~Staub,
  JHEP {\bf 1306}, 073 (2013)
  doi:10.1007/JHEP06(2013)073
  [arXiv:1211.0552 [hep-ph]].

\bibitem{Lu:2013cta}
  X.~Lu, H.~Murayama, J.~T.~Ruderman and K.~Tobioka,
  Phys.\ Rev.\ Lett.\  {\bf 112}, 191803 (2014)
  doi:10.1103/PhysRevLett.112.191803
  [arXiv:1308.0792 [hep-ph]].

\bibitem{Bertuzzo:2014bwa}
  E.~Bertuzzo, C.~Frugiuele, T.~Gregoire and E.~Ponton,
  JHEP {\bf 1504}, 089 (2015)
  doi:10.1007/JHEP04(2015)089
  [arXiv:1402.5432 [hep-ph]].

\bibitem{Staub:2014pca}
  F.~Staub,
  PoS Charged {\bf 2014}, 022 (2014)
  [arXiv:1409.7182 [hep-ph]].

\bibitem{Ding:2015wma}
  R.~Ding, T.~Li, F.~Staub, C.~Tian and B.~Zhu,
  Phys.\ Rev.\ D {\bf 92}, no. 1, 015008 (2015)
  doi:10.1103/PhysRevD.92.015008
  [arXiv:1502.03614 [hep-ph]].

\bibitem{Alvarado:2015yna}
  C.~Alvarado, A.~Delgado, A.~Martin and B.~Ostdiek,
  Phys.\ Rev.\ D {\bf 92}, no. 3, 035009 (2015)
  doi:10.1103/PhysRevD.92.035009
  [arXiv:1504.03683 [hep-ph]].

\bibitem{Bandyopadhyay:2015oga}
  P.~Bandyopadhyay, C.~Coriano and A.~Costantini,
  JHEP {\bf 1509}, 045 (2015)
  doi:10.1007/JHEP09(2015)045
  [arXiv:1506.03634 [hep-ph]].

\bibitem{Evans:2013kxa}
  J.~A.~Evans and D.~Shih,
  JHEP {\bf 1308}, 093 (2013)
  doi:10.1007/JHEP08(2013)093
  [arXiv:1303.0228 [hep-ph]].

\bibitem{Froggatt:1978nt}
  C.~D.~Froggatt and H.~B.~Nielsen,
  Nucl.\ Phys.\  B {\bf 147}, 277 (1979).

\bibitem{Shadmi:2011hs}
  Y.~Shadmi and P.~Z.~Szabo,
  JHEP {\bf 1206}, 124 (2012)
  doi:10.1007/JHEP06(2012)124
  [arXiv:1103.0292 [hep-ph]].

\bibitem{kawa} Y. Kawamura, Prog. Theor. Phys. {\bf 103} (2000) 613;
Prog. Theor. Phys.  {\bf 105}(2001)999; Theor. Phys.  {\bf 105}(2001)691.

\bibitem{GAFF}
G. Altarelli and F. Feruglio,
Phys.\ Lett.\ B {\bf 511}, 257 (2001).


\bibitem{LHYN}
L. Hall and Y. Nomura, Phys.\ Rev.\ D {\bf 64}, 055003 (2001).

\bibitem{AHJMR}
A. Hebecker and J. March-Russell,
Nucl.\ Phys.\ B {\bf 613}, 3 (2001).

\bibitem{Li:2001qs}
  T.~Li,
  Phys.\ Lett.\  B {\bf 520}, 377 (2001);
  Nucl.\ Phys.\  B {\bf 619}, 75 (2001).


\bibitem{Dermisek:2001hp}
R.~Dermisek and A.~Mafi,
Phys.\ Rev.\ D {\bf 65}, 055002 (2002).

\bibitem{Li:2001tx}
T.~Li,
Nucl.\ Phys.\ B {\bf 633}, 83 (2002).


\bibitem{Gogoladze:2003ci}
  I.~Gogoladze, Y.~Mimura and S.~Nandi,
  Phys.\ Lett.\  B {\bf 562}, 307 (2003);
  Phys.\ Rev.\ Lett.\  {\bf 91}, 141801 (2003).





\bibitem{Blumenhagen:2005mu}
  R.~Blumenhagen, M.~Cvetic, P.~Langacker and G.~Shiu,
  Ann.\ Rev.\ Nucl.\ Part.\ Sci.\  {\bf 55}, 71 (2005),
and references therein.

\bibitem{Cvetic:2002pj}
  M.~Cvetic, I.~Papadimitriou and G.~Shiu,
  Nucl.\ Phys.\  B {\bf 659}, 193 (2003)
  [Erratum-ibid.\  B {\bf 696}, 298 (2004)].

\bibitem{Chen:2006ip}
  C.~M.~Chen, T.~Li and D.~V.~Nanopoulos,
  Nucl.\ Phys.\  B {\bf 751}, 260 (2006).





\bibitem{Braun:2005ux}
  V.~Braun, Y.~H.~He, B.~A.~Ovrut and T.~Pantev,
  Phys.\ Lett.\  B {\bf 618}, 252 (2005);
  JHEP {\bf 0605}, 043 (2006),  and references therein.


\bibitem{Bouchard:2005ag}
  V.~Bouchard and R.~Donagi,
  Phys.\ Lett.\  B {\bf 633}, 783 (2006),
  and references therein.





\bibitem{Vafa:1996xn}
  C.~Vafa,
  Nucl.\ Phys.\  B {\bf 469}, 403 (1996).


\bibitem{Donagi:2008ca}
  R.~Donagi and M.~Wijnholt,
  arXiv:0802.2969 [hep-th].

\bibitem{Beasley:2008dc}
  C.~Beasley, J.~J.~Heckman and C.~Vafa,
  JHEP {\bf 0901}, 058 (2009).

\bibitem{Beasley:2008kw}
  C.~Beasley, J.~J.~Heckman and C.~Vafa,
  JHEP {\bf 0901}, 059 (2009).


\bibitem{Donagi:2008kj}
  R.~Donagi and M.~Wijnholt,
  arXiv:0808.2223 [hep-th].




\bibitem{Font:2008id}
  A.~Font and L.~E.~Ibanez,
  JHEP {\bf 0902}, 016 (2009).





\bibitem{Jiang:2009zza}
  J.~Jiang, T.~Li, D.~V.~Nanopoulos and D.~Xie,
  Phys.\ Lett.\  B {\bf 677}, 322 (2009).


\bibitem{Blumenhagen:2008aw}
  R.~Blumenhagen,
  Phys.\ Rev.\ Lett.\  {\bf 102}, 071601 (2009).








\bibitem{Jiang:2009za}
  J.~Jiang, T.~Li, D.~V.~Nanopoulos and D.~Xie,
  Nucl.\ Phys.\  B {\bf 830}, 195 (2010).







\bibitem{Li:2009cy}
  T.~Li,
  Phys.\ Rev.\  {\bf D81}, 065018 (2010)
  [arXiv:0905.4563 [hep-th]].




\bibitem{Li:2010hi}
  T.~Li and D.~V.~Nanopoulos,
  JHEP {\bf 1110}, 090 (2011)
  [arXiv:1005.3798 [hep-ph]].




\bibitem{Kang:2012ra}
  Z.~Kang, T.~Li, T.~Liu, C.~Tong and J.~M.~Yang,
  Phys.\ Rev.\ D {\bf 86}, 095020 (2012)
  [arXiv:1203.2336 [hep-ph]].

\bibitem{Craig:2012xp}
  N.~Craig, S.~Knapen, D.~Shih and Y.~Zhao,
  JHEP {\bf 1303}, 154 (2013)
  [arXiv:1206.4086 [hep-ph]].



\bibitem{MGJS}
M. B. Green and J. H. Schwarz, Phys. Lett. {\bf B149} (1984) 117;
Nucl. Phys. {\bf B255} (1985) 93; M. B. Green, J. H. Schwarz and P.
West, Nucl. Phys. {\bf B254} (1985) 327.


\bibitem{Dreiner:2003hw}
  H.~K.~Dreiner and M.~Thormeier,
  Phys.\ Rev.\ D {\bf 69}, 053002 (2004).

\bibitem{Dreiner:2003yr}
  H.~K.~Dreiner, H.~Murayama and M.~Thormeier,
  Nucl.\ Phys.\ B {\bf 729}, 278 (2005).

\bibitem{Dreiner:2006xw}
  H.~K.~Dreiner, C.~Luhn, H.~Murayama and M.~Thormeier,
  arXiv:hep-ph/0610026.

\bibitem{Gogoladze:2007ck}
  I.~Gogoladze, C.~-A.~Lee, T.~Li and Q.~Shafi,
  Phys.\ Rev.\ D {\bf 78}, 015024 (2008)
  [arXiv:0704.3568 [hep-ph]].

\bibitem{Staub:2008uz}
  F.~Staub,
  arXiv:0806.0538 [hep-ph].

\bibitem{Staub:2009bi}
  F.~Staub,
  Comput.\ Phys.\ Commun.\  {\bf 181}, 1077 (2010)
  doi:10.1016/j.cpc.2010.01.011
  [arXiv:0909.2863 [hep-ph]].

\bibitem{Staub:2010jh}
  F.~Staub,
  Comput.\ Phys.\ Commun.\  {\bf 182}, 808 (2011)
  doi:10.1016/j.cpc.2010.11.030
  [arXiv:1002.0840 [hep-ph]].

\bibitem{Staub:2012pb}
  F.~Staub,
  Comput.\ Phys.\ Commun.\  {\bf 184}, 1792 (2013)
  doi:10.1016/j.cpc.2013.02.019
  [arXiv:1207.0906 [hep-ph]].

\bibitem{Staub:2013tta}
  F.~Staub,
  Comput.\ Phys.\ Commun.\  {\bf 185}, 1773 (2014)
  doi:10.1016/j.cpc.2014.02.018
  [arXiv:1309.7223 [hep-ph]].

\bibitem{Porod:2003um}
  W.~Porod,
  Comput.\ Phys.\ Commun.\  {\bf 153}, 275 (2003)
  doi:10.1016/S0010-4655(03)00222-4
  [hep-ph/0301101].

\bibitem{Porod:2011nf}
  W.~Porod and F.~Staub,
  Comput.\ Phys.\ Commun.\  {\bf 183}, 2458 (2012)
  doi:10.1016/j.cpc.2012.05.021
  [arXiv:1104.1573 [hep-ph]].

\bibitem{Belanger:2013oya}
  G.~Belanger, F.~Boudjema, A.~Pukhov and A.~Semenov,
  Comput.\ Phys.\ Commun.\  {\bf 185}, 960 (2014)
  doi:10.1016/j.cpc.2013.10.016
  [arXiv:1305.0237 [hep-ph]].

\bibitem{Staub:2011dp}
  F.~Staub, T.~Ohl, W.~Porod and C.~Speckner,
  Comput.\ Phys.\ Commun.\  {\bf 183}, 2165 (2012)
  doi:10.1016/j.cpc.2012.04.013
  [arXiv:1109.5147 [hep-ph]].

\bibitem{CMS:2014xfa}
  V.~Khachatryan {\it et al.} [CMS and LHCb Collaborations],
  Nature {\bf 522}, 68 (2015)
  doi:10.1038/nature14474
  [arXiv:1411.4413 [hep-ex]].

\bibitem{Amhis:2014hma}
  Y.~Amhis {\it et al.} [Heavy Flavor Averaging Group (HFAG) Collaboration],
  arXiv:1412.7515 [hep-ex].

\bibitem{ATLAS:2016jaa}
  The ATLAS collaboration [ATLAS Collaboration],
  ATLAS-CONF-2016-077.

\bibitem{CMS:2016inz}
  CMS Collaboration [CMS Collaboration],
  CMS-PAS-SUS-16-030.

\bibitem{Aaboud:2016nwl}
  M.~Aaboud {\it et al.} [ATLAS Collaboration],
  Eur.\ Phys.\ J.\ C {\bf 76}, no. 10, 547 (2016)
  doi:10.1140/epjc/s10052-016-4382-4
  [arXiv:1606.08772 [hep-ex]].

\bibitem{CMS:2016xva}
  CMS Collaboration [CMS Collaboration],
  CMS-PAS-SUS-16-015.

\bibitem{ATLAS:2016kts}
  The ATLAS collaboration [ATLAS Collaboration],
  ATLAS-CONF-2016-078.

\bibitem{CMS:2016inz}
  CMS Collaboration [CMS Collaboration],
  CMS-PAS-SUS-16-030.

\bibitem{Barbieri:1987fn}
  R.~Barbieri and G.~F.~Giudice,
  Nucl.\ Phys.\ B {\bf 306}, 63 (1988).
  doi:10.1016/0550-3213(88)90171-X

\bibitem{Ellis:1986yg}
  J.~R.~Ellis, K.~Enqvist, D.~V.~Nanopoulos and F.~Zwirner,
  Mod.\ Phys.\ Lett.\ A {\bf 1}, 57 (1986).
  doi:10.1142/S0217732386000105




\end{thebibliography}
\end{document}